\documentclass[12pt]{article}

\catcode`\@=11
\@addtoreset{equation}{section}

\global\arraycolsep=2pt
\usepackage{amsbsy,amssymb,latexsym,amsfonts,amsmath}
\usepackage{graphicx,color}

\allowdisplaybreaks

\begin{document}

\begin{flushright}
YITP-23-146
\parbox{4.2cm}

\end{flushright}

\vspace*{0.7cm}

\begin{center}
{ \Large  Central Extension of Scaling Poincar\'e Algebra}
\vspace*{1.5cm}\\
{Yu Nakayama}
\end{center}
\vspace*{1.0cm}
\begin{center}

Yukawa Institute for Theoretical Physics,
Kyoto University, Kitashirakawa Oiwakecho, Sakyo-ku, Kyoto 606-8502, Japan

\vspace{3.8cm}
\end{center}

\begin{abstract}
We discuss the possibility of a central extension of the Poincar\'e algebra and the scaling Poincar\'e algebra.  In more than two space-time dimensions, all the central extensions are trivial and can be removed. In two space-time dimensions, both the Poincar\'e algebra and the scaling Poincar\'e algebra have distinct non-trivial central extensions that cannot be removed. In higher dimensions, the central charges between dilatation and global $U(1)$ symmetry may not be removed.
Based on these central extensions, we give some examples of projective representations of the (scaling) Poincar\'e symmetry in two dimensions.
\end{abstract}

\thispagestyle{empty} 

\setcounter{page}{0}

\newpage

%\section{{Central Extension of Scale Algebra}}

In quantum mechanics, continuous symmetry corresponds to a unitary projective representation of the symmetry group.\footnote{In this note, we do not discuss the non-invertible symmetry which does not define a group, but it may be interesting to study them in the future.} A projective representation of the continuous group can be reduced to an ordinary representation if the two conditions are met: (A) The generators of the group can be redefined so as to remove all central changes from the algebra. (B) The group is simply connected. In this paper, we focus on (A), which is a local constraint (because (B) may be always relaxed by considering the covering group). 
See e.g. \cite{Weinberg:1995mt} for a review. Mathematically, it is equivalent to computing the group cohomology $H^2(G,U(1))$ of the symmetry group $G$ and problem (A) is to determine its non-torsional part.

Consider a central extension of the Lie algebra with generators $t_a$:
\begin{align}
[t_b,t_c] = i C^a_{\ bc} t_a + i C_{bc} \ , 
\end{align}
where $C^a_{\ bc}$ are structure constants and $C_{bc}$ are central charges.  They are constrained by the Jacobi identities:
\begin{align}
C^{a}_{\ bc} C^e_{\ ad} + C^a_{\ c d} C^e_{\ ab} + C^a_{\ db} C^{e}_{\ ac} &= 0 \cr
C^a_{\ bc} C_{ad} + C^{a}_{\ cd} C_{ab} + C^{a}_{\ db} C_{ac}  & = 0 \ .
\end{align}
If there exists a set of real constants $\phi_e$ that satisfies $C_{ab} = C^e_{\ ab} \phi_e$, the central charges can be removed by the redefinition of generators $t_a \to t_a + \phi_a$. 

In a classic paper \cite{Bargmann:1954gh}, Bargmann showed that central charges in (finite) semi-simple algebra can be always removed. The proof is straightforward: by using the Cartan-Killing metric $g_{ab} = C^c_{\ ad} C^{d}_{\ b c}$ available for the semi-simple Lie algebra, one can ``solve" the second Jacobi identity for the central charges as $C_{ab} = - C_a^{\ cd}C_{ed}C^{e}_{\ bc} - C_{a}^{\ cd} C_{ec}C^{e}_{\ d b} = C^{d}_{\ ba} C^c_{\ e} C^{e}_{\ d c}$, where the indices are raised by the (non-degenerate) Cartan-Killing metric, and the first Jacobi identity is used in the last equality. We then see that the central charges can be removed by a redefinition of generators. %he necessity of the semi-simpleness can be easily seen because if we have $U(1)$ generators $t^{U(1)}_i$, the central charges $[t^{U(1)}_i, t^{U(1)}_j] = i C_{ij}^{U(1)}$ cannot be removed. 
The finiteness is necessary because, famously, the Witt algebra generated by $l_n = z^{n+1}\frac{\partial}{\partial z}$, which is semi-simple (but not finite), admits non-trivial central extension (i.e. Virasoro algebra).  

It is of interest to study the possibility of a non-trivial central extension of space-time algebra. In particular, we are interested in if the scaling Poincar\'e algebra (i.e. Poincar\'e and dilatation) admits a non-trivial central extension.\footnote{Note added: the same question in $d=4$ was studied in \cite{Boya:1975uw}, and the conclusion agrees that there is no non-trivial central extension. The paper also studies the structure of the covering group.} First, let us review Weinberg's discussions on the Poincar\'e algebra \cite{Weinberg:1995mt}, which is now generalized in $d$ space-time dimensions.
The central extension of the Poincar\'e algebra is given by
\begin{align}
i[J^{\mu\nu}, J^{\rho\sigma}] &= \eta^{\nu\rho} J^{\mu\sigma} - \eta^{\mu\rho} J^{\nu\sigma} - \eta^{\sigma \mu} J^{\rho\nu} + \eta^{\sigma \nu} J^{\rho \mu} + C^{\rho\sigma,\mu\nu} \cr
i[P^\mu, J^{\rho\sigma}] &= \eta^{\mu\rho} P^\sigma -\eta^{\mu\sigma} P^\rho + C^{\rho\sigma,\mu}  \cr
i[P^\mu, P^\rho] &= C^{\rho, \mu} \ ,
\end{align}
where $J^{\mu\nu} = -J^{\mu\mu}$ are the Lorentz generators and  $P^\mu$ are momentum generators.
The central charges are anti-symmetric
\begin{align}
C^{\rho\sigma, \mu\nu} = -C^{\mu\nu, \rho\sigma} \ ,  \ \ C^{\rho,\mu} = -C^{\mu,\rho} \ . 
\end{align}
Let us study the Jacobi identities for $[J,[P,P]]$, $[J,[J,P]]$ and $[J,[J,J]]$. Of course, we know that the Poincar\'e algebra (without central extensions) is consistent with the Jacobi identities. About central charges, they give non-trivial constraints:
\begin{align}
0 = \eta^{\nu\rho} C^{\mu,\sigma}- \eta^{\mu\rho} C^{\nu,\sigma} - \eta^{\nu\sigma} C^{\mu,\rho} + \eta^{\mu\sigma} C^{\nu,\rho} 
 \label {JPP} \end{align}
\begin{align}
0 =& \eta^{\nu\rho} C^{\mu,\lambda \eta} - \eta^{\mu\rho } C^{\nu,\lambda \eta} - \eta^{\mu\eta} C^{\rho, \lambda \nu} 
 + \eta^{\lambda \mu} C^{\rho, \eta \nu}  \cr 
 &+ \eta^{\lambda \nu} C^{\rho,\mu \eta}- \eta^{\eta nu} C^{\rho,\mu \lambda} + \eta^{\rho\lambda} C^{\eta, \mu\nu} - \eta^{\rho\eta} C^{\lambda, \mu\nu}  \label{JJP}
\end{align}
\begin{align}
0 &= \eta^{\nu\rho} C^{\mu \sigma,\lambda \eta} -  \eta^{\mu\rho} C^{\nu \sigma,\lambda \eta} - \eta^{\sigma \mu } C^{\rho \nu ,\lambda \eta} + \eta^{\sigma \nu } C^{\rho \mu ,\lambda \eta}  \cr
&+\eta^{\eta \mu } C^{\lambda \nu,\rho \sigma} -\eta^{\lambda \mu } C^{\eta \nu,\rho \sigma} - \eta^{\nu \lambda } C^{\mu \eta,\rho \sigma} + \eta^{\nu \eta} C^{\mu \lambda ,\rho \sigma} \cr
& + \eta^{\sigma \lambda } C^{\rho \eta ,\mu \nu} -  \eta^{\rho \lambda } C^{\sigma \eta ,\mu \nu} -  \eta^{\eta \rho } C^{\lambda \sigma ,\mu \nu} +  \eta^{\eta \sigma} C^{\lambda \rho,\mu \nu} \ . 
\label{JJJ}
\end{align}

Contracting \eqref{JPP} with $\eta_{\nu\rho}$, we have
\begin{align}
(d-2) C^{\mu,\sigma} = 0 \ .
\end{align}
Unless $d=2$, it means $C^{\mu,\sigma}=0$. For now, let us assume $d>2$ and later consider the case $d=2$ separately.

Contracting \eqref{JJP} with $\eta_{\nu\rho}$, we obtain
\begin{align}
C^{\mu,\lambda \eta} = \eta^{\mu\eta} C^\lambda - \eta^{\mu\lambda} C^\eta
\end{align}
Here $C^\lambda = \frac{1}{d-1} \eta_{\rho\nu} C^{\rho, \lambda \nu}$. (There is no Lorentz generator in $d=1$, so we assumed $d\neq 1$.) Contracting \eqref{JJJ} with $\eta_{\nu\rho}$, we obtain
\begin{align} 
C^{\mu \sigma,\lambda \eta} = \eta^{\eta \mu}C^{\lambda \sigma } -\eta^{\lambda \mu} C^{\eta \sigma} + \eta^{\sigma \lambda} C^{\eta \mu} - \eta^{\eta \sigma} C^{\lambda \mu} \ . 
\end{align}
Here $C^{\lambda \sigma} = \frac{1}{d-2} \eta_{\rho\nu} C^{\lambda \nu, \sigma \rho}$. 

We now show that unless $d=2$, the central extension of the Poincar\'e algebra is trivial. Indeed they can be removed by the redefinition of the generators
\begin{align}
\tilde{P}^\mu &= P^{\mu} + C^\mu \cr
\tilde{J}^{\mu\sigma}  &= J^{\mu\sigma} + C^{\mu\sigma} \ . 
\end{align}
The new generators with tilde satisfy the same Poincar\'e algebra without central extensions. (Physically it is just a constant shift of the energy-momentum and angular-momentum).

In $d=2$, we have seen $C^{\mu,\nu}$ can be non-zero. Accordingly, there does exist a non-trivial central extension of the Poincar\'e algebra:
\begin{align}
i[P^0, P^1] &= C_P \cr
i[J^{01}, P^0] & = P^1 \cr
i[J^{01}, P^1] & = -P^0 \ .  \label{cL}
\end{align}
More covariantly, we may put $\epsilon^{01}=1$ in front of $C_P$. The existence of the second-rank epsilon tensor makes $d=2$ distinct. The centrally extended algebra is solvable. Physically, this central extension is equivalent to introducing space-time non-commutativity (in momentum space). An example is shown at the end of this note (see example 1), but it is worthwhile mentioning that the algebra is sometimes called the ``Maxwell algebra" because it appears in the motion of a charged particle in a constant electric field.
We will later see that this central extension is not compatible with the additional dilatation symmetry. 

Going back to general dimensions, let us add dilatation to the Poincar\'e algebra. Including the central extensions, the additional commutation relations are
\begin{align}
i[D,P^\mu] &= P^\mu + c^\mu \cr 
i[D,J^{\mu\nu}] &= c^{\mu\nu} \ . 
\end{align}
Here, $c^\mu$ and $c^{\mu\nu}$ are potential new central charges.
For now, let us assume $d\neq 2$ and the Poincar\'e part of the generators are chosen so that the central charges discussed above (i.e. $C^{\mu,\nu}, C^{\mu,\lambda \eta}, C^{\mu\nu,\lambda\eta}$) are absent. We will then show that $c^\mu$ and $c^{\mu\nu}$ are incompatible with the Jacobi identities (unless $d=2$, in which case we will have a separate discussion)

To see this, let us consider the Jacobi identities of $[D,[J,P]$ and $[D,[J,J]]$. The central charges must satisfy  the conditions
\begin{align}
0 &= \eta^{\mu\rho} c^\sigma - \eta^{\mu\sigma} c^\rho \cr
0 &=  \eta^{\nu\rho} c^{\mu\sigma} - \eta^{\mu\rho} c^{\nu\sigma} - \eta^{\sigma \mu} c^{\rho\nu} + \eta^{\sigma \nu} c^{\rho \mu} \  . \label{Jacobid}
\end{align}
Contracting the first equation of \eqref{Jacobid} with $\eta_{\mu\rho}$, we have
\begin{align}
0 = (d-1) c^\sigma \ . 
\end{align}
Contracting the second equation of \eqref{Jacobid} with $\eta_{\nu\rho}$ then, we have
\begin{align}
0 = (d-2) c^{\mu \sigma} \ . 
\end{align}
Thus, unless $d=1,2$, a non-trivial central extension of the scale algebra does not exist. In $d=1$, the Jacobi identities allow a central extension of the scale algebra: $i[P^0,D] = P^0 + c^0$, but it can be removed by shifting $P^0$. Note that since there is no Lorentz generator in $d=1$, we have not fixed the shift of $P^0$ yet, and we can use the freedom here.

Let us focus on the exceptional case of $d=2$. As promised, let us first show that the central charge in the two-dimensional Poincar\'e algebra $C_P$ in \eqref{cL} is not consistent with the dilatation. This is obtained by studying $[D,[P^0,P^1]$ Jacobi identity, leading to $C_P=0$. More physically, $C_P$ would be dimensionful so it could not be consistent with the dilatation.
In contrast, the existence of  $c^{01}$ is compatible with all the Jacobi identities and cannot be removed in $d=2$. Thus we have the non-trivial central extension of the scaling Poincar\'e algebra in $d=2$: 
\begin{align}
i[P^0, P^1] &= 0 \cr
i[J^{01}, P^0] & = P^1 \cr
i[J^{01}, P^1] & = -P^0 \cr
i[D,P^0] &= P^0 \cr
i[D,P^1] &= P^1 \cr
i[D,J^{01}] & = c^{01} \ .\label{cd}
\end{align}
More covariantly, one may replace $c^{01}$ with $\epsilon^{01} c_J$. The centrally extended algebra is solvable.  Physically, the angular momentum has an anomalous scaling dimension and the dilatation has an anomalous angular momentum,
An example is shown at the end of this note (see example 2). 

In the passing, let us mention that (global) conformal algebra $SO(d-2,2)$ is semi-simple, so there is no non-trivial central extension by Bargmann's theorem \cite{Bargmann:1954gh} alluded to at the beginning. In two dimensions, there is another exotic space-time symmetry that is compatible with relativity: Poincar\'e + scale + left-moving conformal (without right-moving conformal). Bargemann's theorem tells that left-moving part (i.e. $SO(2,1)$, which is semi-simple) does not admit any non-trivial central extension while it is also easy to see that the central extension of the right-moving part $[\bar{L}_{-1}, \bar{L}_0] = -\bar{L}_{-1} + \bar{c}$ can be removed by a  shift of $\bar{L}_{-1}$.
We also note that central extensions of the superalgebra in various dimensions are known to exist. For example, $\mathcal{N}=2$ supersymmetry in $d=4$ admits the non-trivial central charges: $\{Q^1_\alpha, Q^2_\beta \} = Z \epsilon_{\alpha \beta}$.

Let us briefly discuss the addition of global symmetries. Repeating the analysis, we can see that due to the Lorentz symmetry, there is no non-trivial central extension between Poincar\'e (or conformal) generators and the global symmetry generators unless in $d=2$, where $i[J^{\mu\nu}, Q] = c_{JQ} \epsilon^{\mu\nu}$ is possible with $Q$ being a global $U(1)$ charge.\footnote{Of course, $U(1)$ global generators may have non-trivial central extensions among themselves.}  In a theory with dilatation (without special conformal transformation) in any dimension, the non-trivial central extension $i[D,Q] = c_Q$ is possible.\footnote{When $Q$ has a non-trivial scale dimension, whose possibility was overlooked in the classic paper \cite{Haag:1974qh}, we could have $i[D,Q] = \Delta_Q Q + c_Q$ in principle (under the absence of the special conformal transformation). In the case $\Delta_Q \neq 0$, we can remove the central charge $c_Q$ by shifting $Q$.  Note, however, that the shift of $Q$ may introduce additional central charges among global symmetries.  The non-zero $\Delta_Q$ with no $c_Q$ appear in interacting shift-symmetric field theories such as the ones studied in \cite{Gimenez-Grau:2023lpz}.} It leads to a logarithmic theory where the dilatation cannot be diagonalized and the existence of the Noether current for $Q$ may be questionable. See example 3 below.

Some examples are in order. They form projective representations of the Poincar\'e algebra (with dilatation). The examples below are free, and it would be interesting to see if there are any interacting ones.

Example 1. Consider a left-moving particle in $1+1$ dimensions with momentum $p_+$. It is described by the wavefunction $\psi(p_+)$. Let us define the right-moving momentum operator by $p_- = i \frac{\partial}{\partial p_+}$, then it has the centrally extended Poincar\'e algebra $i[P_+,P_-] = C_P$, where the central charge is the particle number. We can also check that the representation of $p_-$ by the derivative of $p_+$ is consistent with the Lorentz transformation but not with the dilatation. 

This construction suggests that if there is a one-dimensional particle described by a canonical variable $(x,p)$ with the Hamiltonian $H= x + p$, it is a projective representation of the two-dimensional Poincar\'e symmetry. It has a ``Lorentz generator" $J = xp + px$, but no dilatation generator.

This model has a familiar analogy to a falling apple (or a charged particle in a constant electric field). 
There we consider a non-relativistic particle under the constant gravitational field with the Hamiltonian $H= \frac{p_z^2}{2m} + mgz$. We usually say that the translational symmetry is broken by the constant gravitational field. Indeed, the Lagrangian or Hamiltonian is not invariant under the translation in $z$. However, we also see that the equation of motion is invariant under the translation. There is no conserved charge in the usual sense, but the translation symmetry is realized in a projective manner: $i[H,p_z] = mg$. In the path integral language, one may also say that the translational symmetry is anomalous because the action is shifted by a constant under the translation. As such, it may be useful in constraining possible renormalization group flow.

Example 2. Consider a massless particle in $1+1$ dimensions with internal degrees of freedom represented by  $q \in \mathbb{R}$. We define the dilatation by $D = P^+ \frac{\partial}{\partial P^+} +  P^- \frac{\partial}{\partial P^-}   + q$ and the Lorentz transformation by $J =  P^+ \frac{\partial}{\partial P^+} -  P^- \frac{\partial}{\partial P^-}   + i c_J \frac{d}{dq}$. It realizes the centrally extended scaling Poincar\'e algebra \eqref{cd}. 

Since there is no projective representation of the (global) conformal algebra due to Bargmann's theorem, the projective representation of the scaling Poincar\'e symmetry cannot be uplifted to the (global) conformal symmetry. About the
 Zamolodchikov-Polchinski theorem \cite{Zamolodchikov:1986gt}\cite{Polchinski:1987dy} (see \cite{Nakayama:2013is} for review) that dilatation invariance implies (global) conformal invariance in two-dimensional relativistic theories, this model violates the assumption of the discrete spectrum due to the infinite internal degrees of freedom represented by $q$. Fourier transform of $q$ may be regarded as a ``continuous spin".\footnote{Of course, without interactions, we can always define dilatation and Lorentz transformation without a contribution from $q$ and then this non-centrally extended algebra can be embedded in the conformal algebra. Thus, the more relevant question here is whether we can introduce interactions while preserving the continuous spin degrees of freedom.}

Example 3. Consider a massless scalar particle in any dimension with internal degrees of freedom represented by $q \in \mathbb{R}$. We define the dilatation by $D= p^\mu \frac{\partial}{\partial p^\mu} + i c_Q \frac{\partial}{\partial q}$, and the global $U(1)$ charge $Q =  q$. It realizes the centrally extended algebra $i[D,Q] = c_Q$. Since it is not consistent with $[P^\mu, K^\nu] = 2i J^{\mu\nu} + 2i \eta^{\mu\nu} D$, we have to dispose of the special conformal generators $K^\mu$.

The projective representation may be regarded as an anomaly in quantum mechanics: a projective representation cannot be continuously deformed to be an ordinary representation without breaking the symmetry or closing energy gaps. It would be interesting to see if some of these structures studied in this paper may be useful to classify space-time symmetry in physical systems. In the example of a falling apple (example 1 above), we can imagine that there is a phase transition of the energy spectrum at $g=0$, which may be regarded as a consequence of the anomaly.

Finally, let us note that the centrally-extended Poincar\'e algebra has attracted some interest in the context of flat space holography (e.g. \cite{Afshar:2019axx}\cite{Godet:2020xpk}\cite{Engelhardt:2020qpv}\cite{Grumiller:2020elf}\cite{Ruzziconi:2020wrb}\cite{Afshar:2021qvi}\cite{Kar:2022sdc}\cite{Rosso:2022tsv} and reference therein), where the gravitational realization was introduced by Cangemi and Jackiw\footnote{I had several email communications with Roman (e.g. about the existence of linearly realized scale invariant field theories with non-linearly realized conformal symmetry), but I think I had only one chance of in-person discussion with him at Texas in 2013, where I have a fond memory of enjoying Texas BBQ together with stimulating conversation.} sometime ago \cite{Cangemi:1992bj}\cite{Cangemi:1993sd}. It would be interesting to see if the centrally-extended scaling Poincar\'e algebra may have similar holographic applications. However, we should point out that unlike the centrally extended Poincar\'e algebra, the centrally extended scaling Poincare\'e algebra does not admit any non-degenerate invariant bilinear form because all operators have non-negative grading with respect to $D$. The existence of the non-degenerate invariant bilinear form was a crucial element in writing down the BF-like action in \cite{Cangemi:1992bj}\cite{Cangemi:1993sd}.

This work is in part supported by JSPS KAKENHI Grant Number 21K03581. The author would like to thank J.~ F.~Carinena for informing him of the work \cite{Boya:1975uw}.


\begin{thebibliography}{99}


%\cite{Weinberg:1995mt}
\bibitem{Weinberg:1995mt}
S.~Weinberg,``The Quantum theory of fields. Vol. 1: Foundations,''
Cambridge University Press, 2005,
ISBN 978-0-521-67053-1, 978-0-511-25204-4
doi:10.1017/CBO9781139644167
%617 citations counted in INSPIRE as of 18 Sep 2023




%\cite{Bargmann:1954gh}
\bibitem{Bargmann:1954gh}
V.~Bargmann,
``On Unitary ray representations of continuous groups,''
Annals Math. \textbf{59}, 1-46 (1954)
doi:10.2307/1969831
%415 citations counted in INSPIRE as of 18 Sep 2023

%\cite{Boya:1975uw}
\bibitem{Boya:1975uw}
L.~J.~Boya, J.~F.~Carinena and M.~Santander,
%``Dilatations and the Poincare Group,''
J. Math. Phys. \textbf{16}, 1813-1815 (1975)
doi:10.1063/1.522756
%7 citations counted in INSPIRE as of 20 Nov 2023


%\cite{Haag:1974qh}
\bibitem{Haag:1974qh}
R.~Haag, J.~T.~Lopuszanski and M.~Sohnius,
%``All Possible Generators of Supersymmetries of the s Matrix,''
Nucl. Phys. B \textbf{88}, 257 (1975)
doi:10.1016/0550-3213(75)90279-5
%1582 citations counted in INSPIRE as of 18 Oct 2023

%\cite{Gimenez-Grau:2023lpz}
\bibitem{Gimenez-Grau:2023lpz}
A.~Gimenez-Grau, Y.~Nakayama and S.~Rychkov,
%``Scale without Conformal Invariance in Dipolar Ferromagnets,''
[arXiv:2309.02514 [hep-th]].
%2 citations counted in INSPIRE as of 18 Oct 2023


%\cite{Zamolodchikov:1986gt}
\bibitem{Zamolodchikov:1986gt}
A.~B.~Zamolodchikov,
``Irreversibility of the Flux of the Renormalization Group in a 2D Field Theory,''
JETP Lett. \textbf{43}, 730-732 (1986)
%1620 citations counted in INSPIRE as of 18 Sep 2023


%\cite{Polchinski:1987dy}
\bibitem{Polchinski:1987dy}
J.~Polchinski,
``Scale and Conformal Invariance in Quantum Field Theory,''
Nucl. Phys. B \textbf{303}, 226-236 (1988)
doi:10.1016/0550-3213(88)90179-4
%442 citations counted in INSPIRE as of 18 Sep 2023

%\cite{Nakayama:2013is}
\bibitem{Nakayama:2013is}
Y.~Nakayama,
``Scale invariance vs conformal invariance,''
Phys. Rept. \textbf{569}, 1-93 (2015)
doi:10.1016/j.physrep.2014.12.003
[arXiv:1302.0884 [hep-th]].
%313 citations counted in INSPIRE as of 18 Sep 2023



%\cite{Afshar:2019axx}
\bibitem{Afshar:2019axx}
H.~Afshar, H.~A.~Gonz\'alez, D.~Grumiller and D.~Vassilevich,
%``Flat space holography and the complex Sachdev-Ye-Kitaev model,''
Phys. Rev. D \textbf{101}, no.8, 086024 (2020)
doi:10.1103/PhysRevD.101.086024
[arXiv:1911.05739 [hep-th]].
%49 citations counted in INSPIRE as of 15 Nov 2023



%\cite{Godet:2020xpk}
\bibitem{Godet:2020xpk}
V.~Godet and C.~Marteau,
%``New boundary conditions for AdS$_{2}$,''
JHEP \textbf{12}, 020 (2020)
doi:10.1007/JHEP12(2020)020
[arXiv:2005.08999 [hep-th]].
%34 citations counted in INSPIRE as of 15 Nov 2023



%\cite{Engelhardt:2020qpv}
\bibitem{Engelhardt:2020qpv}
N.~Engelhardt, S.~Fischetti and A.~Maloney,
%``Free energy from replica wormholes,''
Phys. Rev. D \textbf{103}, no.4, 046021 (2021)
doi:10.1103/PhysRevD.103.046021
[arXiv:2007.07444 [hep-th]].
%75 citations counted in INSPIRE as of 15 Nov 2023

%\cite{Grumiller:2020elf}
\bibitem{Grumiller:2020elf}
D.~Grumiller, J.~Hartong, S.~Prohazka and J.~Salzer,
%``Limits of JT gravity,''
JHEP \textbf{02}, 134 (2021)
doi:10.1007/JHEP02(2021)134
[arXiv:2011.13870 [hep-th]].
%46 citations counted in INSPIRE as of 15 Nov 2023


%\cite{Ruzziconi:2020wrb}
\bibitem{Ruzziconi:2020wrb}
R.~Ruzziconi and C.~Zwikel,
%``Conservation and Integrability in Lower-Dimensional Gravity,''
JHEP \textbf{04}, 034 (2021)
doi:10.1007/JHEP04(2021)034
[arXiv:2012.03961 [hep-th]].
%40 citations counted in INSPIRE as of 15 Nov 2023


%\cite{Afshar:2021qvi}
\bibitem{Afshar:2021qvi}
H.~Afshar and B.~Oblak,
%``Flat JT gravity and the BMS-Schwarzian,''
JHEP \textbf{11}, 172 (2022)
doi:10.1007/JHEP11(2022)172
[arXiv:2112.14609 [hep-th]].
%12 citations counted in INSPIRE as of 15 Nov 2023


%\cite{Kar:2022sdc}
\bibitem{Kar:2022sdc}
A.~Kar, L.~Lamprou, C.~Marteau and F.~Rosso,
%``A Matrix Model for Flat Space Quantum Gravity,''
JHEP \textbf{03}, 249 (2023)
doi:10.1007/JHEP03(2023)249
[arXiv:2208.05974 [hep-th]].
%10 citations counted in INSPIRE as of 15 Nov 2023

%\cite{Rosso:2022tsv}
\bibitem{Rosso:2022tsv}
F.~Rosso,
%``A solvable model of flat space holography,''
JHEP \textbf{02}, 037 (2023)
doi:10.1007/JHEP02(2023)037
[arXiv:2209.14372 [hep-th]].
%7 citations counted in INSPIRE as of 15 Nov 2023


%\cite{Cangemi:1992bj}
\bibitem{Cangemi:1992bj}
D.~Cangemi and R.~Jackiw,
%``Gauge invariant formulations of lineal gravity,''
Phys. Rev. Lett. \textbf{69}, 233-236 (1992)
doi:10.1103/PhysRevLett.69.233
[arXiv:hep-th/9203056 [hep-th]].
%191 citations counted in INSPIRE as of 15 Nov 2023

%\cite{Cangemi:1993sd}
\bibitem{Cangemi:1993sd}
D.~Cangemi and R.~Jackiw,
%``Poincare gauge theory for gravitational forces in (1+1)-dimensions,''
Annals Phys. \textbf{225}, 229-263 (1993)
doi:10.1006/aphy.1993.1058
[arXiv:hep-th/9302026 [hep-th]].
%59 citations counted in INSPIRE as of 15 Nov 2023



\end{thebibliography}
\end{document}